# Information Storage and Retrieval using Macromolecules as Storage Media

M. Mansuripur[†], P.K. Khulbe[†], S.M. Kuebler[‡], J.W. Perry[‡†], M.S. Giridhar[†],
J. Kevin Erwin[†], Kibyung Seong[†], Seth Marder[‡†], and N. Peyghambarian[†]

[†]Optical Sciences Center and [‡]Department of Chemistry, University of Arizona,
Tucson, Arizona, 85721 masud@u.arizona.edu

**Introduction**. To store information at extremely high-density and data-rate, we propose to adapt, integrate, and extend the techniques developed by chemists and molecular biologists for the purpose of manipulating biological and other macromolecules. In principle, volumetric densities in excess of $10^{21}$ bits/cm$^3$ can be achieved when individual molecules having dimensions below a nanometer or so are used to encode the 0's and 1's of a binary string of data. In practice, however, given the limitations of electron-beam lithography, thin film deposition and patterning technologies, molecular manipulation in submicron dimensions, etc., we believe that volumetric storage densities on the order of $10^{16}$ bits/cm$^3$ (i.e., petabytes per cubic centimeter) should be readily attainable, leaving plenty of room for future growth. The unique feature of the proposed new approach is its focus on the feasibility of storing bits of information in individual molecules, each only a few angstroms in size. Nature provides proof of principle for this type of data storage through the ubiquitous existence of DNA and RNA molecules, which encode the blueprint of life in four nucleic acids: Adenine, Guanine, Cytosine, and Thymine/Uracil.[1] These macromolecules are created on an individual basis by enzymes and protein-based machinery of biological cells, are stable over a fairly wide range of temperatures, are read and decoded by ribosomes (for the purpose of manufacturing proteins), and can be readily copied and stored under normal conditions.

Advances in molecular biology over the past decades have made it possible to create (i.e., write) artificial molecules of arbitrary base-sequence, and also to decode (i.e., read) such sequences. These techniques can now be adapted and extended in the service of a new generation of ultra-high-density data storage devices. Macromolecular strings (representing blocks of data several megabytes long) can be created on-demand, then stored in secure locations (parking spots) on a chip. These data blocks can then be retrieved by physically moving them to decoding stations (also located on the same chip), subjecting them to a "read" process, then returning them to their secure parking spots until the next request for readout is issued, or until there is a call for their removal and destruction (i.e., erasure).

**System architecture**. The required parking lots as well as the read, write, and erase stations for data encoding/decoding can be fabricated in integrated fashion on the surface of a glass substrate using lithography or other surface patterning techniques. Transfer of the molecular data blocks between the parking lots and the various read/write/erase stations may be achieved via controlled electric-field gradients, optical tweezers, micro-fluidic pumps, opto-electronic micro-motors, etc. Schemes for high-resolution reading and writing of data blocks utilizing electro-photo-chemical processes will be described in the following sections. As for achievable data rates, although the individual read/write stations envisioned in this paper may not be able to handle data rates in excess of a few megabits per second, the possibility of using multiple parallel read/write stations should enable the proposed scheme to compete with the projected data-transfer-rates in conventional optical and magnetic storage technologies.

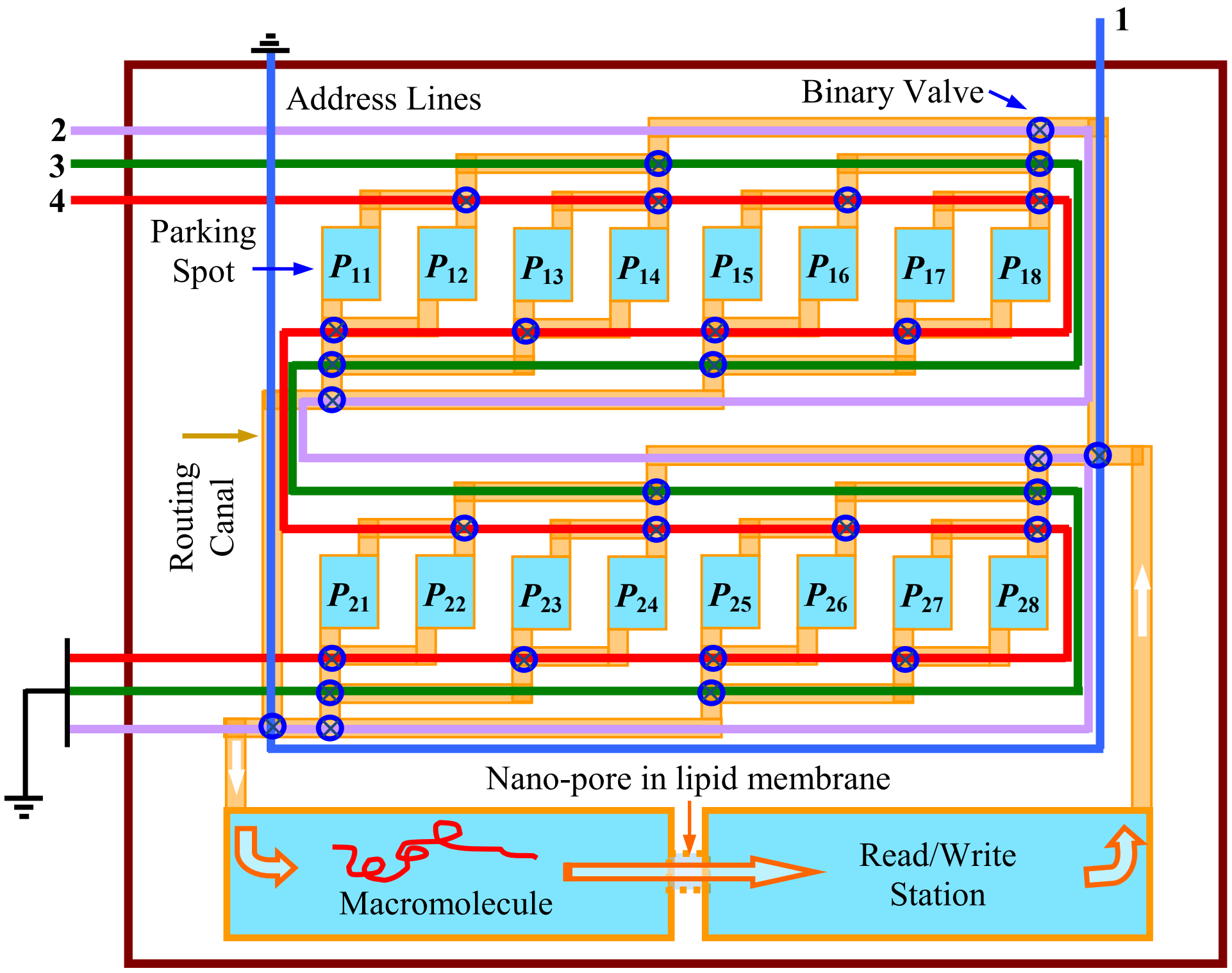

**Fig.1**. Diagram showing the patterned surface of the proposed data storage chip. Once a parking spot is selected, its macromolecular content will be transferred to the read/write station under the influence of an applied electric voltage (i.e., electro-phoretic transfer). Following the completion of the read/write operation, the macromolecule is returned to its designated spot.

A possible implementation of the proposed storage device, depicted schematically in Fig. 1, shows a specific arrangement of 16 parking spots in conjunction with a single read/write station on a chip surface. Various canals are etched on this surface to connect the parking spots to the read/write chamber. Binary valves placed at the cross-sections of these canals control the flow direction (a possible design of a binary μ-valve appears in Fig. 2). Controlling the valves is accomplished by means of electrical signals on four separate lines.

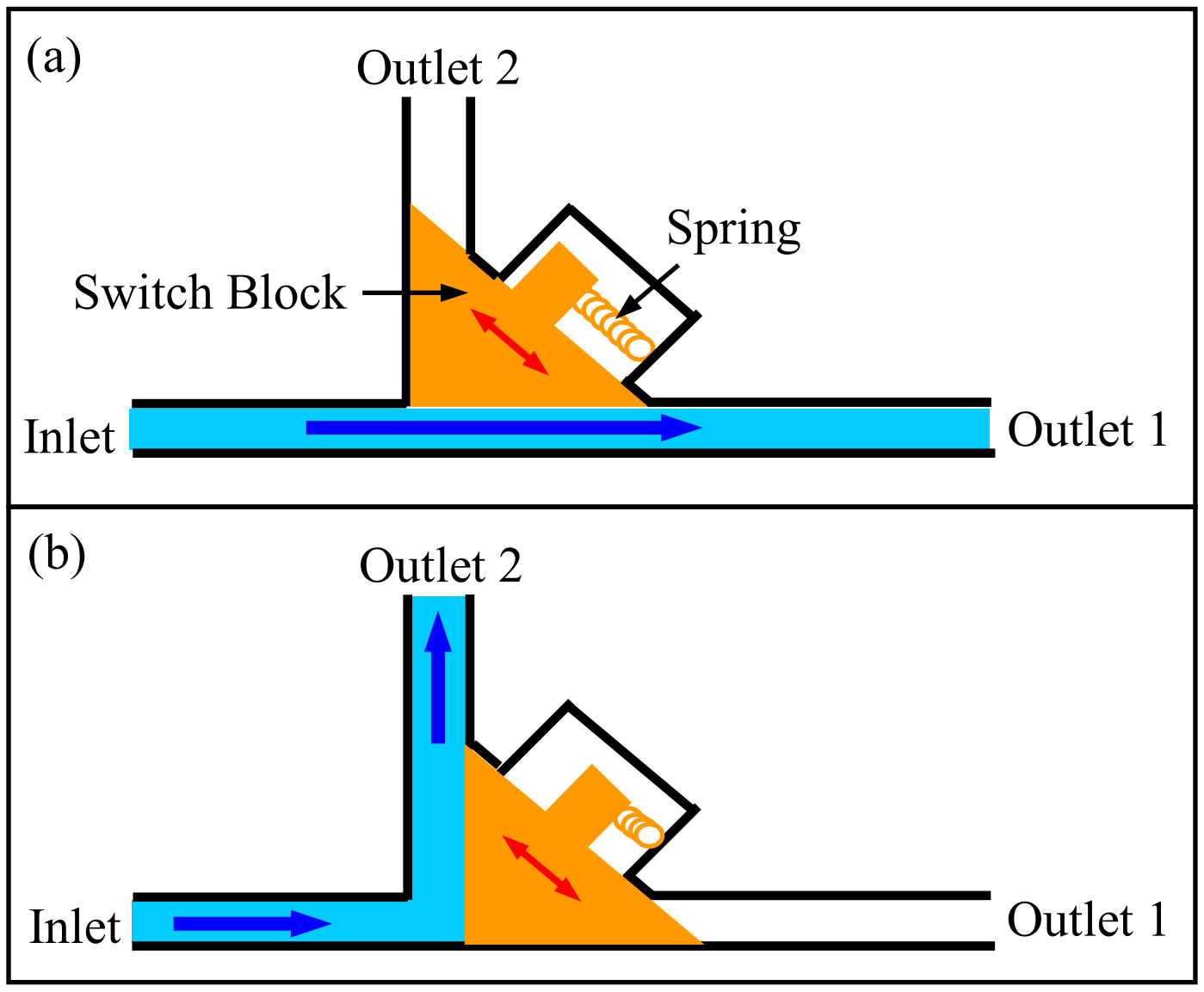

**Fig.2**. Binary valve channels the flow of incoming liquid toward either of outlets 1 or 2. An electronic command signal is required to compress the spring and pull the switch block to the lower position. When the command signal is removed, the spring relaxes to its initial state, thus redirecting the flow toward outlet 1.

The aforementioned binary scheme of grouping the parking spots results in a number of address lines that is the base-2 logarithm of the number of spots. One can thus address over a million parking spots on the surface of a chip with only 20 lines. If each parking spot, for example, occupies a $10 \times 10 \mu m^2$ area, one million such spots will cover a square centimeter of the chip surface. With the macromolecule stored in each parking spot corresponding to a mega-byte block of user data, the areal density of storage on the chip will approach 1 $TeraBytes/cm^2$. Moreover, if the parking spots (and associated read/write/erase stations) are stacked in 10μm-thick layers, the volumetric capacity of the envisioned device will easily reach 1 PetaBytes/$cm^3$.

**Macromolecular readout by translocation through a nano-pore**. The diagrams in Fig. 3 depict a method of readout by translocation through a nano-pore.[2] As different nucleotides pass through the pore, they hamper the flow of the ionic current differently. Fluctuations in current blockage are due to differences in the size and/or charge of the various nucleotides. Given a sufficient SNR, one should be able to uniquely identify the base sequence of the translocated DNA molecule by analyzing the electrolytic current waveform.

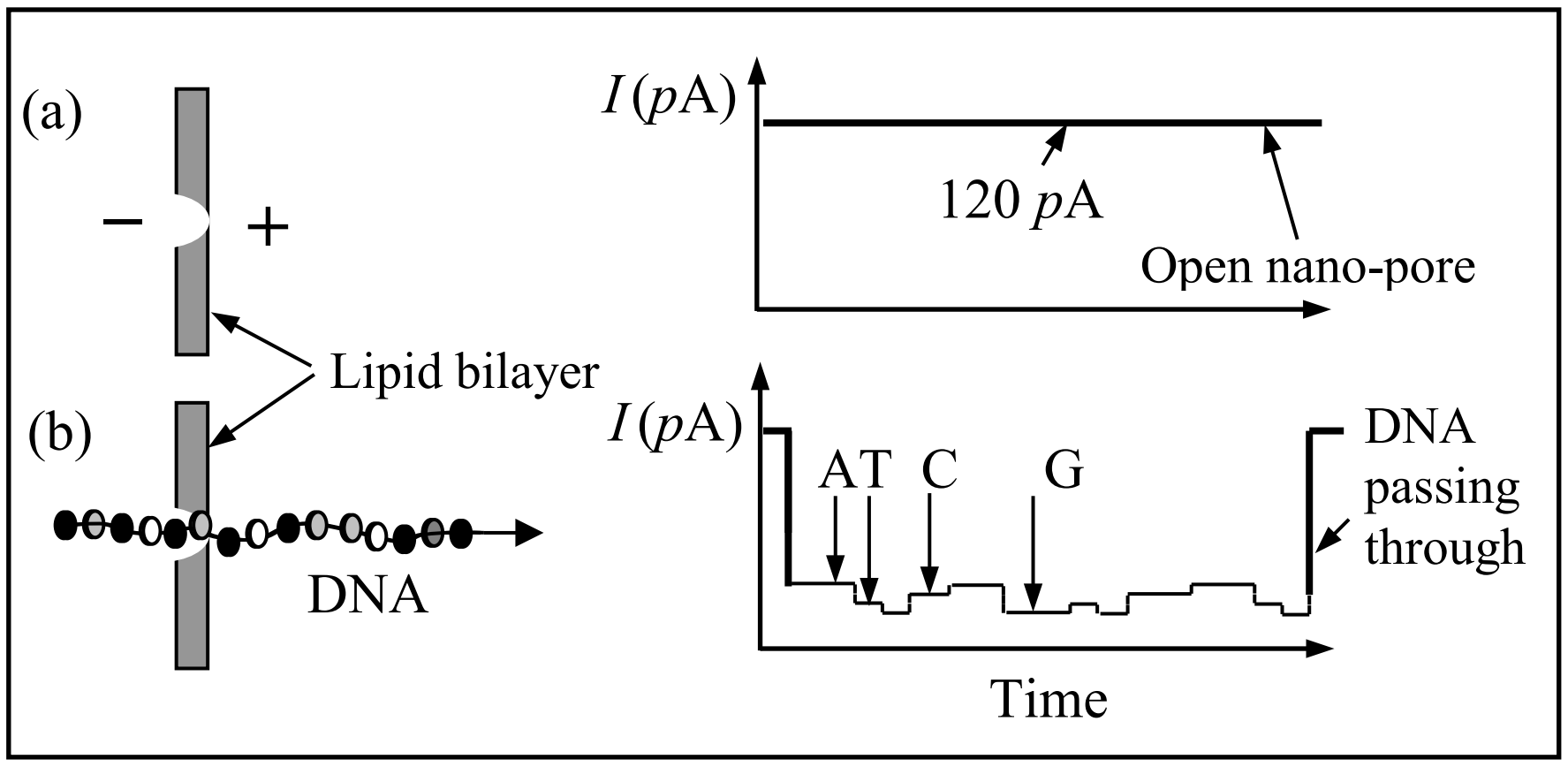

**Fig. 3.** (a) Nano-pore created in a lipid membrane by α-hemolysin proteins. The bilayer separates two sections of a buffer solution. With 100 mV applied across the bilayer, ~120 pA of ionic current flows through the nano-pore. (b) DNA strand passing through the nano-pore partially blocks the ionic current.

Figure 4 is a schematic diagram of a setup for conducting DNA translocation experiments in our laboratory. The system consists of two ~150μL wells, machined in a teflon block and connected by a U-shaped tube. One end of the U-tube has a sleeve of teflon tubing, which is rapidly tapered to provide a ~ 40μm diameter aperture. Ag-AgCl electrodes are inserted into both wells. To form a lipid bilayer at the 40μm aperture, one must first prime the aperture by applying 5μL of a 200μg/mL solution of lipid (diphytanoyl phosphatidylcholine) dissolved in spectroscopic-grade hexane. The primer is then allowed to evaporate in a mild stream of nitrogen gas. The wells on both sides of the aperture are subsequently filled with a buffer solution (1.0M KCl + 10 mM HEPES/KOH, pH = 8.0). At this point, a relatively large ionic current passes through the aperture (several hundred nA at 120mV applied voltage). To form a lipid bilayer, a single bristle of a fine brush is dipped in a ~3mg/mL lipid solution (dissolved in spectroscopic-grade hexadecane), then brushed across the aperture. Formation of the bilayer is indicated by the blockage of the ionic current (several hundred GΩ resistance between the Ag-AgCl electrodes).

Once a stable bilayer has been created, 80 ng of α-hemolysin in 2μL buffer solution is added to the well that contains the aperture and, subsequently, a 120mV potential is applied to the electrodes. Typically, a single ion-channel consisting of seven α-hemolysin molecules self-assembles in a time span of 5-20 minutes; successful self-assembly is indicated by an abrupt increase of the electrolytic current to 120pA. To avoid formation of multiple ion-channels, the excess α-hemolysin is removed by perfusion with ~20mL of fresh buffer. (In our experiments, individual ion-channels were stable for up to six hours.) With an ion-channel formed stably within the lipid bilayer, we add single-stranded DNA molecules (ssDNA) to the cis (i.e., grounded) side of the bilayer. Ionic current through the α-hemolysin channel is then monitored using a HEKA EPC-9 patch-clamp amplifier in voltage clamp mode. Data were acquired at 5μs intervals and filtered with a low-pass Bessel filter (bandwidth = 5, 20, or 100kHz). The passage of DNA strands through the ion-channel is indicated by transient blockage events in the ionic current. Several hundred blockage events were recorded and analyzed for poly-A, poly-C, and poly-AC samples.

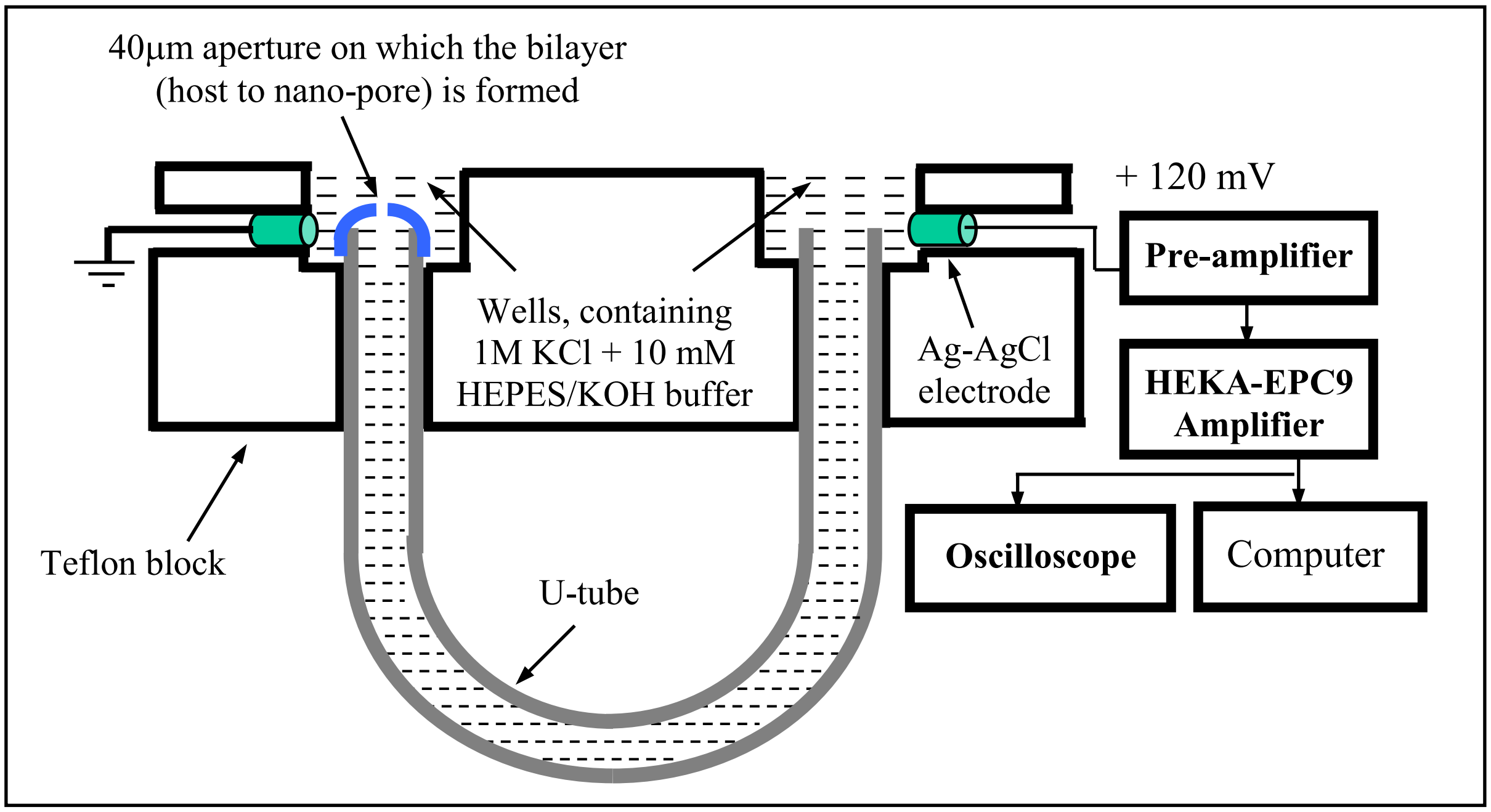

**Figure 4**. Experimental setup for translocation studies. Two 150μL wells are machined in a teflon block and connected via a U-shaped teflon tube (inner diameter ~1.6 mm). The electrodes, which have a silver core with AgCl coating, are connected to a constant voltage source.

**Translocation experiments conducted with poly-A, poly-C, and poly-AC**. Size-purified ssDNA samples were purchased from *Midland Certified Reagents* (Midland, TX). In a typical experiment these DNA samples were added to the cis chamber (concentration ~200 nM/mL). About one hundred translocation events were recorded separately for each type of DNA sample. Figure 5(a) shows three typical translocation events of a ssDNA molecule having 120 continuous adenine bases (A-120). Similarly, Fig. 5(b) shows translocation events associated with DNA molecules having 120 cytosine bases (C-120). A comparison of these events reveals that adenine bases block the ionic current more than the cytosine bases. During translocation of A-120, the ionic current drops from 130*p*A to 20*p*A, whereas in the case of C-120 the current drops from 128*p*A to 40*p*A. The experiments also reveal that cytosine bases translocate faster than the adenine bases. The translocation speed for C-120 is ~1.5μs/base, while that for A-120 is either ~3.3μs/base or ~5.4μs/base, depending on the direction of translocation (5'→3' or 3'→5'). It is thus seen that 120-base poly-A and poly-C molecules of ssDNA can be readily distinguished from each other by measuring the magnitude and/or duration of the current blockage.

To determine whether smaller segments of A-bases can be distinguished from similar segments of C-bases, we conducted experiments with a 120-base sample having the periodic sequence 20A-20C-20A-20C-20A-20C. The results, depicted in Fig. 5(c), indicate that alternate 20A and 20C segments cannot be distinguished via current blockage measurements. This is because the length of the α-hemolysin ion-channel in the translocation direction is ~10nm, while each DNA base is only 0.34nm long. At any given time, therefore, there are roughly 30 bases in the channel, which explains why 20-base-long sequences are irresolvable with this particular ion-channel.

For a single DNA base, the transit time through the ion-channel can be as short as 1.5μs (poly-C). To resolve alternate segments, each consisting of ~30 identical bases, one should monitor the ionic current with a fairly wide-band amplifier. (The highest bandwidth presently available to us is 100kHz.) One must also try to substantially reduce the noise, whose magnitude, as present measurements indicate, rises in proportion to the area of the lipid bilayer. Reduction of the buffer volume (by reducing the chamber size) should also be an effective way of reducing the noise.

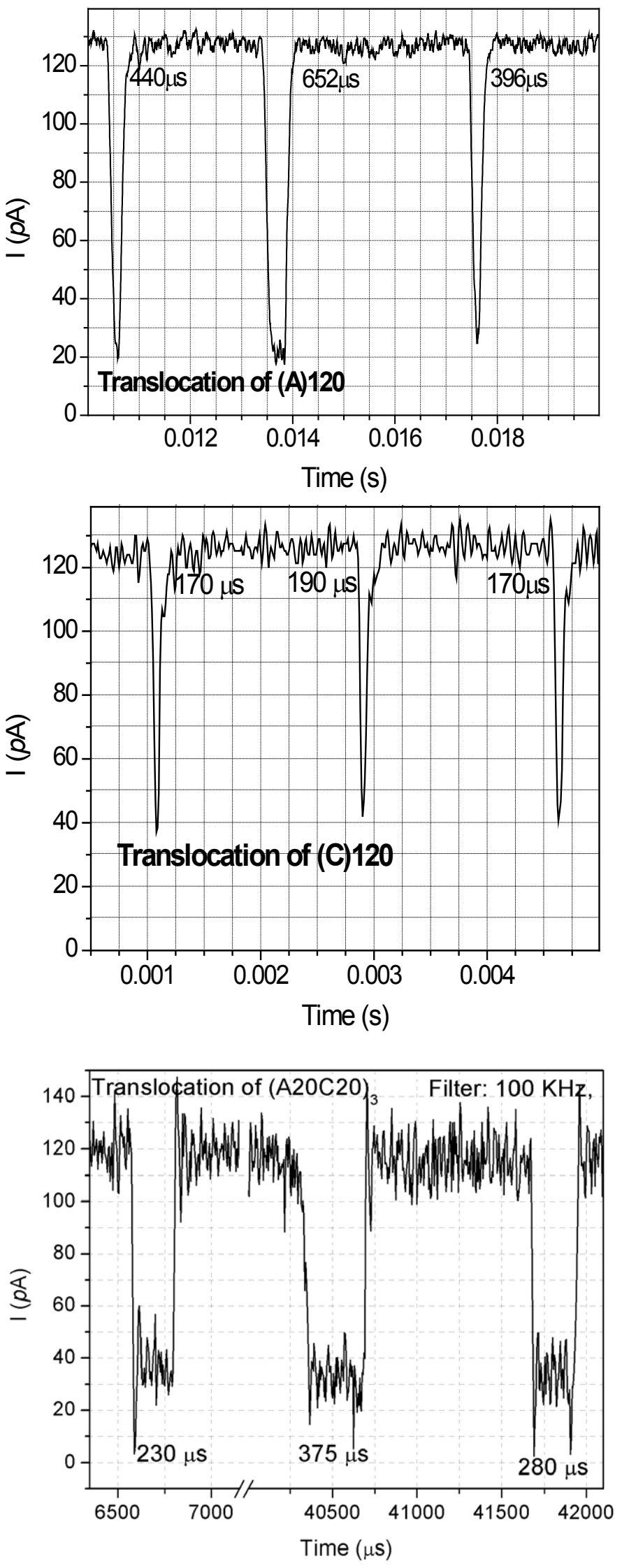

**Figure 5.** Three typical current blockade events during translocation of purified, 120-base-long ssDNA molecules. (a) With poly-A strands, the ionic current drops from ~130 pA to ~20 pA, with a blockage duration of either ~ 400μs or ~ 650 μs, depending on the direction of travel of the molecule. (b) With poly-C strands, the current drops from ~128 pA to ~ 40 pA; the blockade duration per cytosine base is ~ 1.5μs. (c) With ssDNA having 20 bases of adenine alternating with 20 bases of cytosine the current drops from ~120 pA to ~35 pA during translocation. (The poly-AC data is recorded at a larger amplifier bandwidth, which results in a higher noise level compared to the poly-A and poly-C data.)

**Micro-chamber fabrication**. To realize high storage density and high data-rate, the system architecture requires miniaturized parking lots for storing macromolecular strings (i.e., data blocks) as well as integrated read/write/erase stations for data encoding and decoding. We have fabricated a prototype read station on a microscope slide using 2-photon-initiated polymerization of certain resins.[3] A 150μm-thick photo-polymerizable resin is coated on a glass slide, upon which the desired pattern is written by tracing a highly focused laser beam. The localized photo-excitation cross-links the resin, thereby reducing the solubility of the exposed material, and the desired 3D structure is obtained by washing away the unexposed resin. Our read station made by this method consists of two adjacent chambers, $150 \times 150 \times 150\mu m^3$, separated by a partition wall, as shown in Fig. 6. The partition wall has a 20μm diameter hole at its center, upon which a lipid bilayer must be formed to host the nano-pore. This and similar structures will be used for DNA translocation experiments in the next phase of our program. Figure 7 shows the water-filling and subsequent drying of the chambers. The first frame of Fig. 7 shows a micro-pipette, in contact with the partition wall, injecting water into one of the chambers. The third frame shows the over-filled section of the chamber with a convex surface (water meniscus). As the water evaporates, the meniscus first becomes flat at $t = 14$s, then concave at $t = 22$s. The chamber eventually dries up at $t = 31$s.

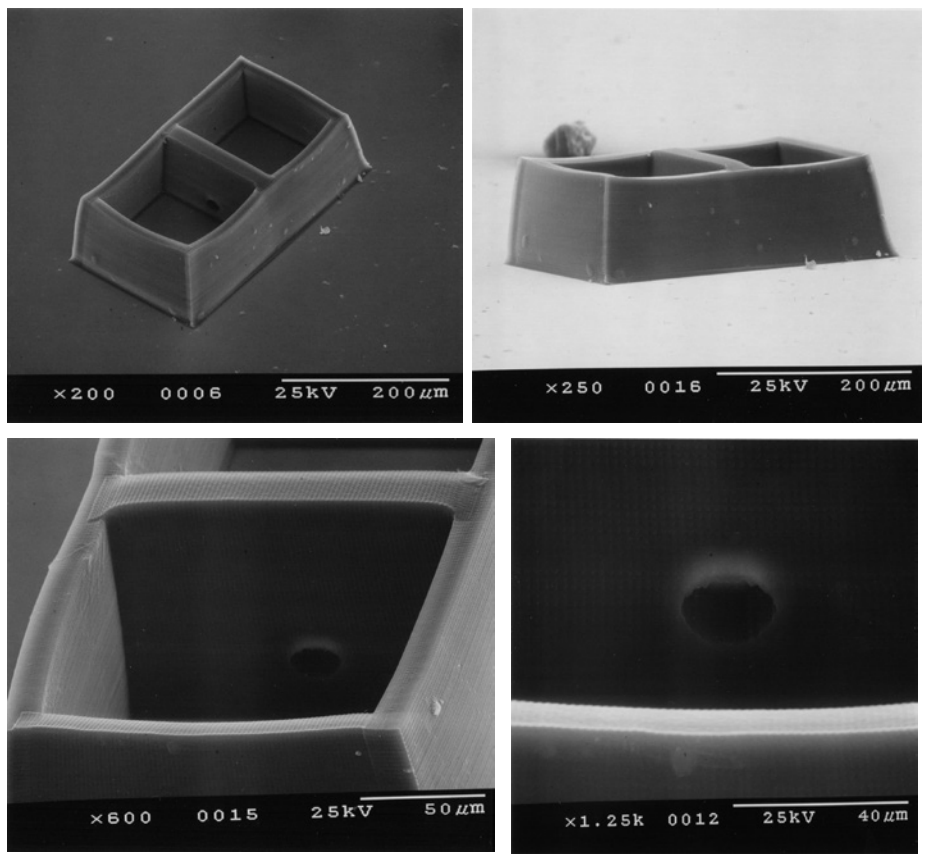

**Figure 6**. (Top) Two views of the micro-chamber fabricated by 2-photon lithography. Each chamber is approximately 150μm on the side, and about 150μm deep. (Bottom) Two views of the μ-hole in the wall separating the two chambers. The μ-hole is ~20 μm in diameter and is located ~20μm above the chambers' floor.

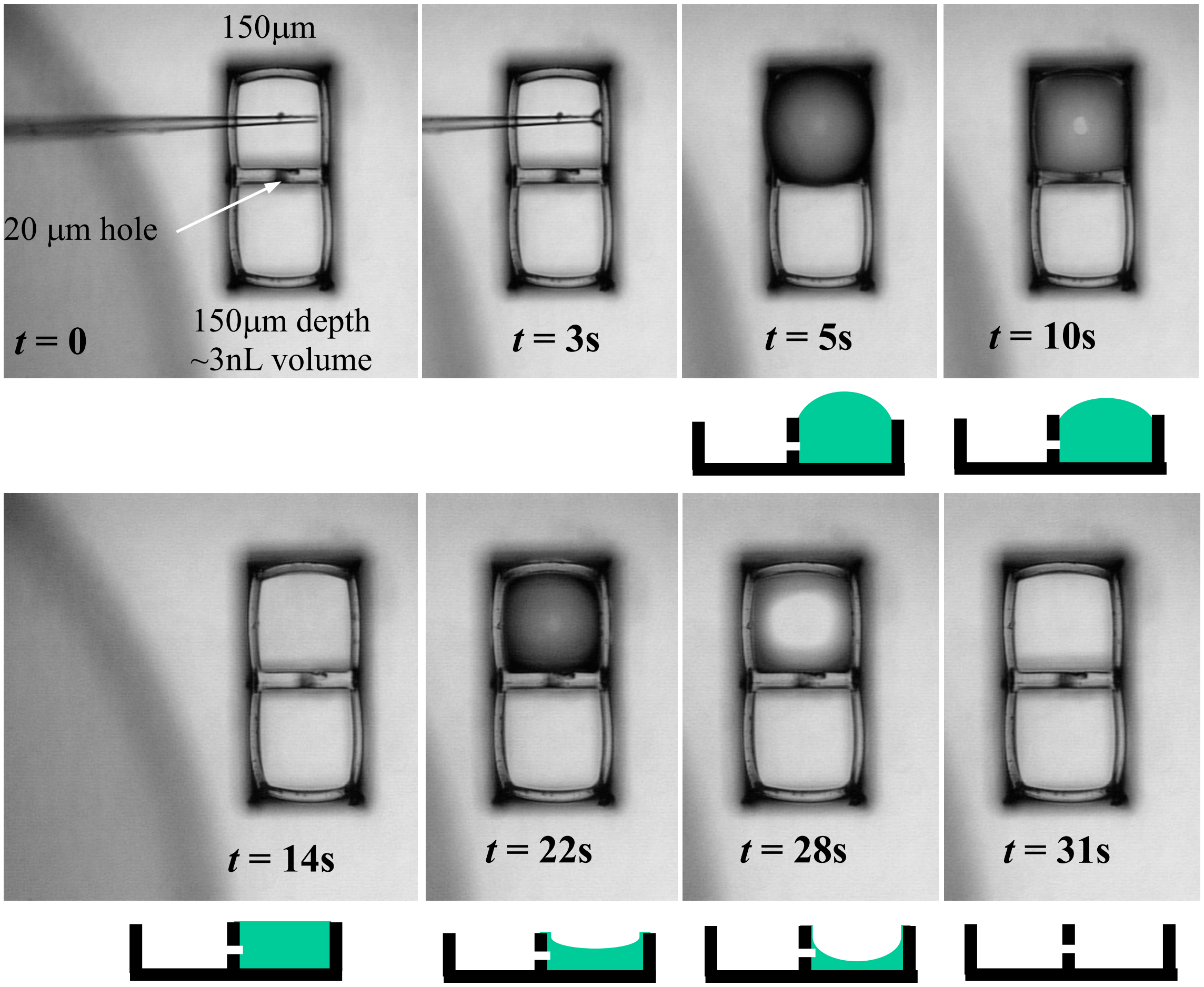

**Figure 7**. Water filling sequence of a μ-chamber. A μ-pipette (inner diameter ~10μm) is brought in contact with a side-wall to release its content. Each μ-chamber can hold ~3nL of water, which completely evaporates in less than 30 seconds at the room temperature under normal conditions.

To prevent evaporation, we fabricated covered (i.e., roofed) chambers having access holes on the end walls to enable μ-pipette insertion. One such device is shown in Fig. 8, where the μ-pipette is seen to be inserted through the wall on the left-hand side. The injected liquid goes through the central hole and starts leaking out of the hole on the right-hand wall. (The water continues to leak until the pressure inside the pipette falls to the atmospheric level.) The leaked droplets can be sucked back into the chambers, as shown in the last frame of Fig. 8.

The above μ-chambers' shortcomings for use as DNA read stations include: (i) high evaporation rates, (ii) structural fragility that renders them unusable after only a few experiments, and (iii) difficulty of attaching electrodes, performing perfusion, or accessing the central hole to paint the lipid bilayer and/or to inject the nano-pore proteins.

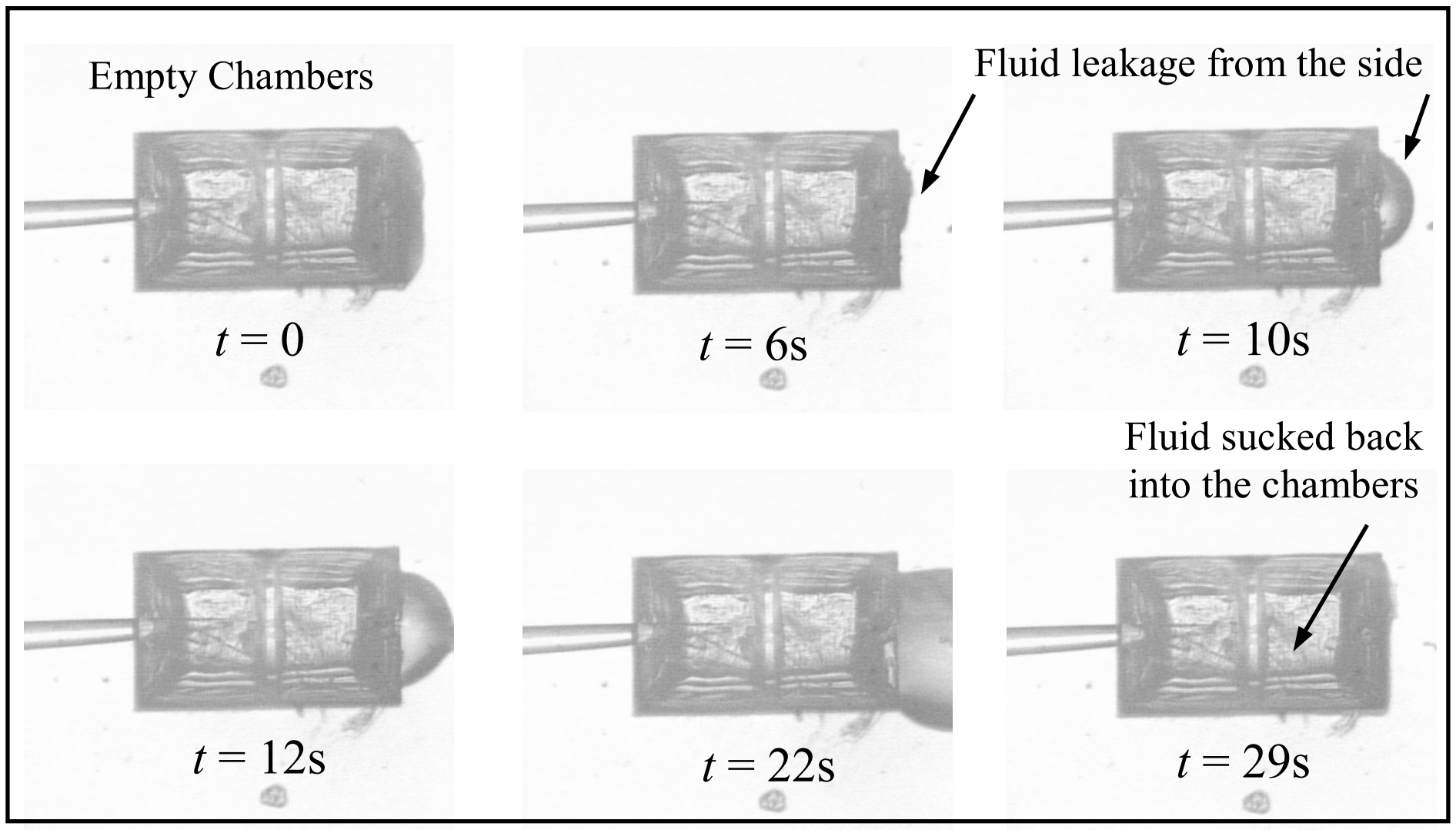

**Figure 8**. Filling a covered μ-chamber with water via a μ-pipette inserted through a hole on a side-wall. The last frame at $t$ = 29s shows the liquid being sucked back into the chambers.

**Device fabrication using polydimethylsiloxane (PDMS)**. To integrate the read station with the parking spots, to reduce the noise level during readout (which noise arises in part from the large chamber size as well as the large surface area of the lipid membrane), and in order to overcome the aforementioned shortcomings of the 2-photon micro-fabrication method, we attempted to miniaturize the read station of Fig. 1 using several alternative techniques. In one such attempt, we built a device by molding a PDMS block, as can be seen in Fig. 9. The liquid PDMS precursor was poured into a cup, and thin tungsten wires were placed in the liquid at locations where it was desired to create channels, chambers, and access ports. The cup was then heated to ~50°C, where it was kept for two to three hours until the PDMS solidified. Subsequently the tungsten wires were removed, the device was fitted with Ag-AgCl electrodes, and the chambers were filled with a buffer solution via μ-pipettes. In Fig. 9 the read station has two cylindrical chambers connected by a tapered aperture (diameter ~150μm) over which a lipid bilayer will be formed. There are two ports for Ag-AgCl electrodes, and a perfusion line to flush out the excess α-hemolysin after ion-channel formation. The inset to Fig. 9 is a photograph showing the central section of the fabricated device.

Capillary pipettes used for filling and emptying the above chambers were made by a commercial pipette-puller (Narishige Scientific Instruments), which has a circular filament that heats the center of a glass capillary to the glass softening temperature. Both ends of the capillary are held by iron rods and pulled by an adjustable magnetic force. The length of the pipette taper is controlled by adjusting the filament current (heating) and the magnetic force. The capillaries used in the above experiments have an inner tip diameter in the range of 10 to 50μm; the taper length is in the 1-2cm range.

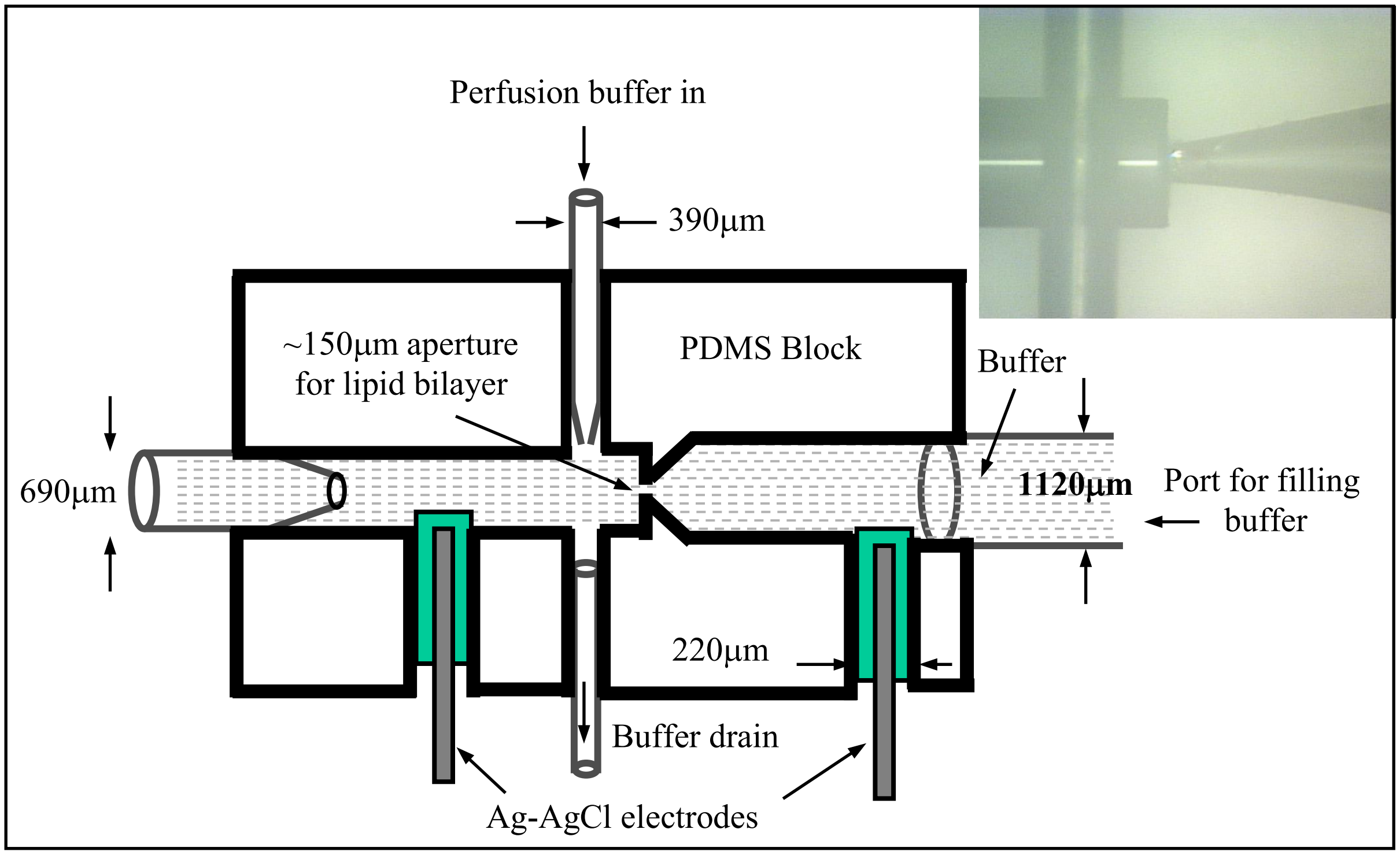

**Figure 9.** Miniature DNA read station made by molding a PDMS block. The inset is a photograph of the central region of the fabricated read station.

**Femto-second pulse laser micro-machining of a glass substrate**. An alternative method of fabricating the μ-channels and μ-chambers of the read station is based on femto-second pulse laser micro-machining of a glass substrate. Figure 10 shows scanning electron micrographs of a polished glass slide on which three chambers, each ~$180 \times 180 \times 100\mu m^3$, have been written. One of these chambers is used for calibration purposes, while the other two are connected via a μ-hole at the center of the partition wall. The micro-machining was done with a (time-averaged) laser power of 18mW, wavelength = 1.66μm, pulse width = 150 fs, repetition rate = 1 kHz, focusing lens NA = 0.25. For the chambers the scanning speed was 50μm/s, with 4μm steps in the $XY$-plane and 15μm steps in the depth direction $Z$. The conical hole at the center of the partition wall has a diameter ~ 40μm on the front side and ~10μm on the rear side. The hole was created by focusing the laser beam through the edge of the glass slide onto the partition wall that separates the chambers. Creating the hole required a scanning speed in the range of 1-5μm/s, with 2μm steps in the $X$, $Y$, and $Z$-directions. In practice, these μ-chambers must be covered with a solid PDMS layer to prevent the evaporation of liquids.

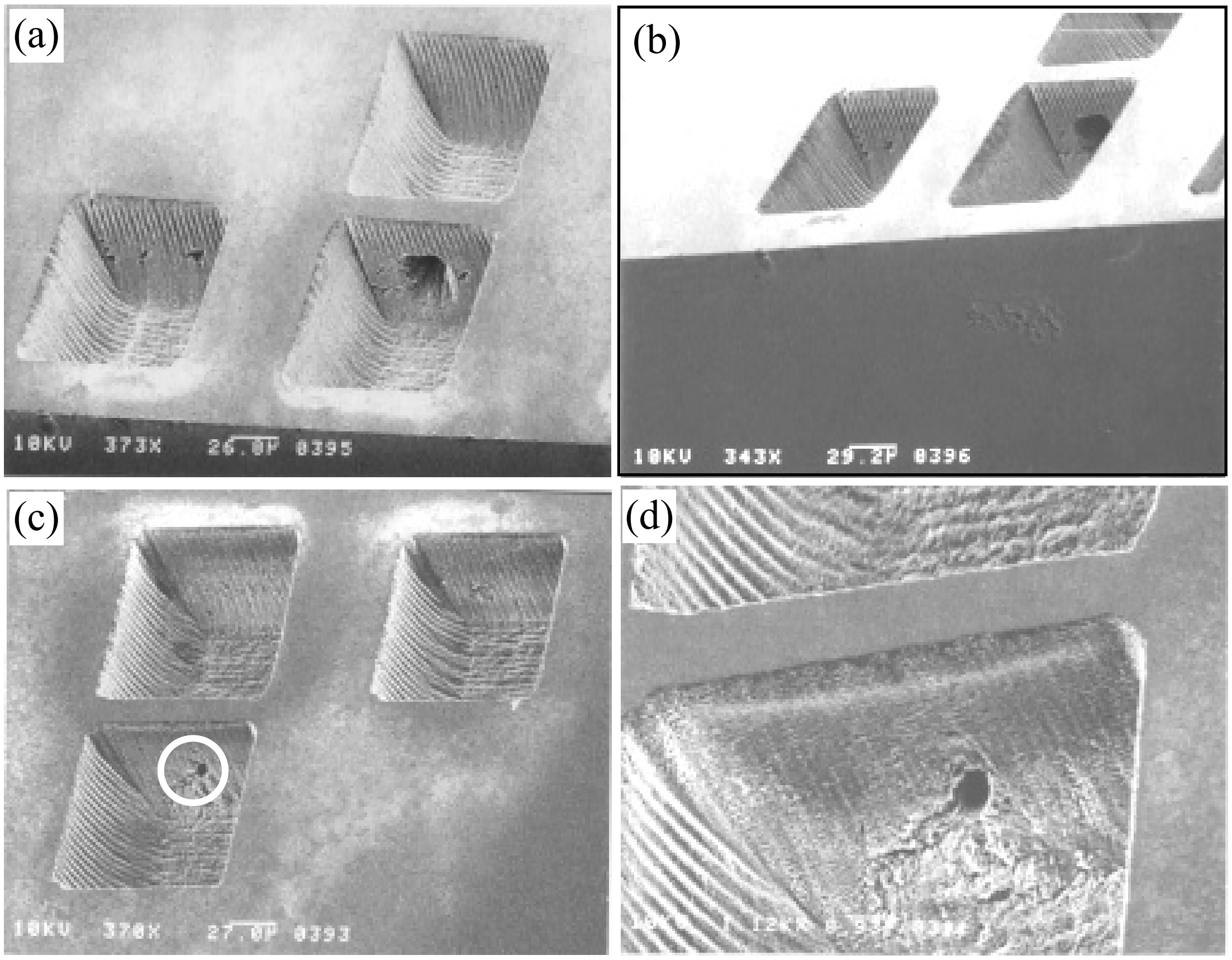

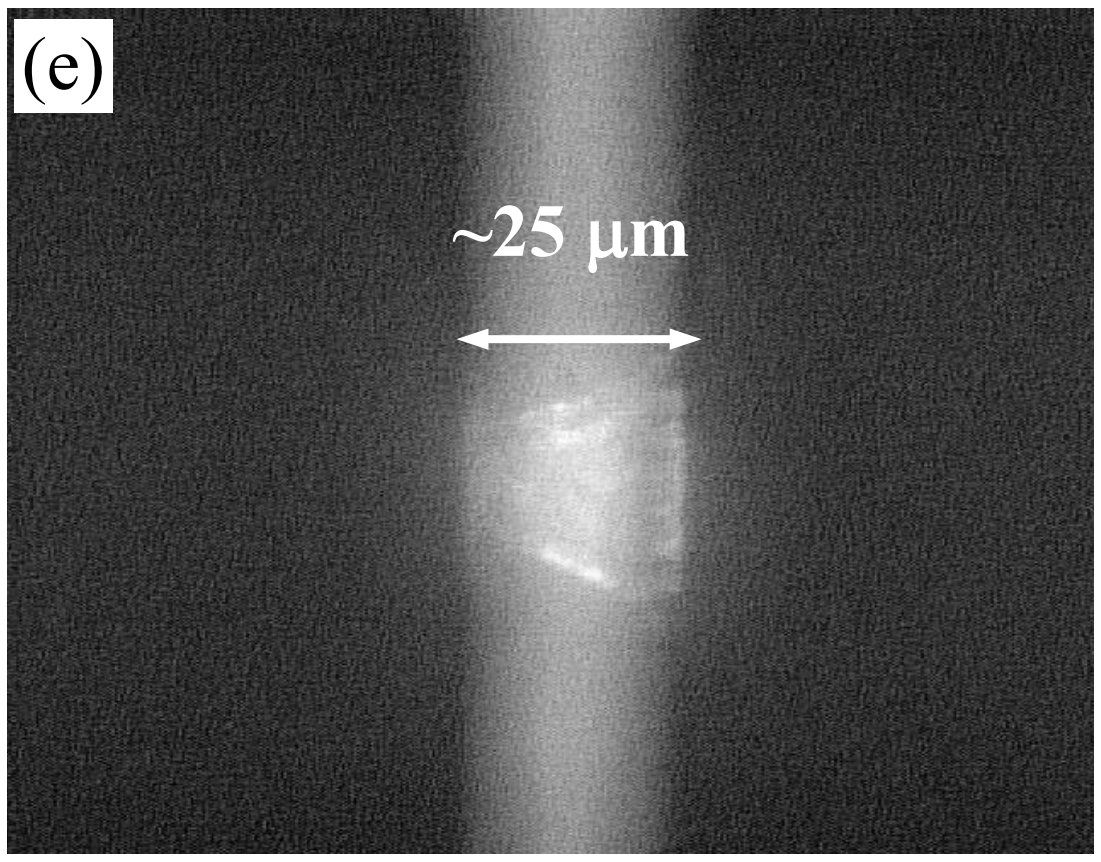

**Figure 10**. Femtosecond pulse laser micro-machining of a glass substrate. Of the three chambers shown, one is for calibration purposes, while the other two provide a miniaturized read station. (a, b) Front view of the μ-hole, showing the wide-end of the cone-shaped aperture having ~ 40μm diameter. (c) Rear view of the hole, where the aperture diameter at its narrow end is only ~10μm. (d) Magnified view of the hole as seen from the rear. (e) Optical micrograph of the hole as seen from the top of the partition wall.

**Micro-pump action using a focused, circularly-polarized laser beam**. Fluid circulation by means of a micro-pump is one of the envisioned methods of transporting macro-molecular data blocks between the parking spots and the read/write stations. Anticipating the need for such micro-pumps, we have investigated methods of circulating micro-fluids under focused laser beams. The ultimate goal is to fabricate these micro-pumps in an integrated fashion around the periphery of the chip, where they can be accessed and controlled by external laser beams focused through transparent side-windows. Figure 11 shows a birefringent particle submerged in water and rotating under a focused, circularly-polarized beam of light. Birefringent particles, ranging in size from 1μm to 10μm, were produced by grinding a half-wave plate. The particles were suspended in a 150μm-thick layer of water, and circularly-polarized laser light ($\lambda$ = 635nm) was focused onto individual particles via a 0.6NA microscope objective lens. The observed rotation of the particles was due to the transfer of angular momentum from circularly-polarized photons. On varying the laser power in the range from 3.0 mW to 9.0 mW, we observed uniform rotation of the particles with rotational frequencies ranging from 0.8 Hz to 5.0 Hz. These particles could be rotated in both clockwise and counterclockwise directions, depending on whether the focused beam was right- or left-circularly polarized. We believe the combination of micro-fluidic techniques with remote pump control via an external laser beam is a powerful tool for macro-molecular transport within a micro-fluidic chip.

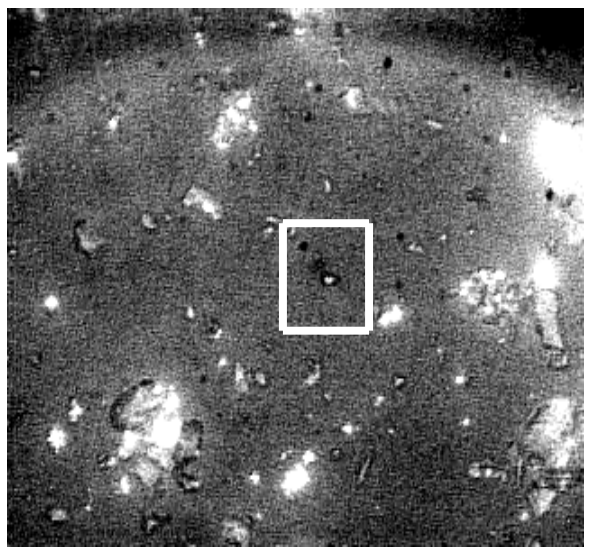

**Figure 11.** Birefringent particle (diameter ~ 5μm) rotates under a focused, circularly-polarized laser beam. The photograph shows a large number of particles (obtained by grinding a half-wave plate) suspended in water and observed through an optical microscope within its field of view. The microscope objective (NA = 0.6) is used to focus a red laser beam (λ = 635nm) onto the particle seen here near the center of the frame. Laser beam scattering from the spinning particle creates a rotating beacon that is readily visible to the naked eye.

**Writing individual macromolecules**. A method of writing macromolecules of arbitrary base sequence is shown in Fig. 12. This is essentially a miniaturized and automated version of the photo-chemical technique used in Gene-Chip manufacture.[4] A short strand of a precursor molecule is needed to initiate the recording process, whereby individual bases are sequentially added to the free end of the precursor. The precursor is anchored at the center of the write station to prevent its drift. The write station is illuminated by a focused laser beam for the purpose of removing the blocking molecule attached to the growing end of the macromolecule. The various bases (i.e., A, C, G, T in the case of deoxyribonucleic acids) are kept within specific reservoirs and blocked by a chemical group that prevents the growth of the molecule beyond a single base during each visit to the reservoir. The reservoirs are connected to the main chamber of the write station through a nano-pore; opening each nano-pore (e.g., by applying a proper voltage between the reservoir and the main chamber) causes the free-floating end of the macromolecule to enter the selected reservoir. Within the reservoir a single base attaches itself to the free end of the growing macromolecule, but further growth will be prevented by the presence of the blocking group. The macromolecule is subsequently returned to the main chamber, where the laser beam removes its blocking group in preparation for entry into the next reservoir. This writing scheme, although probably too slow for practical purposes, has the advantage that its chemistry is well understood and, therefore, can provide a proof of principle for the proposed method of molecular data storage.

**Alternative writing scheme based on widely-separated active molecules**. Another possible writing mechanism is depicted in Fig. 13. In (a) a long precursor strand consisting of relatively long, inert segments separating individual active molecules, which are identical in their native (or ground) state, is exposed to an activator. The inert segments simply act as spacers to create a reasonable distance between adjacent active molecules. The activator may be a tightly focused laser beam, a localized electric field, a tunneling electrode, an anchored enzyme, etc. The strand in Fig. 13(a) is being pulled under the activator by a micro-machine (not shown); for instance, optical tweezers may be dragging the strand from right to left.

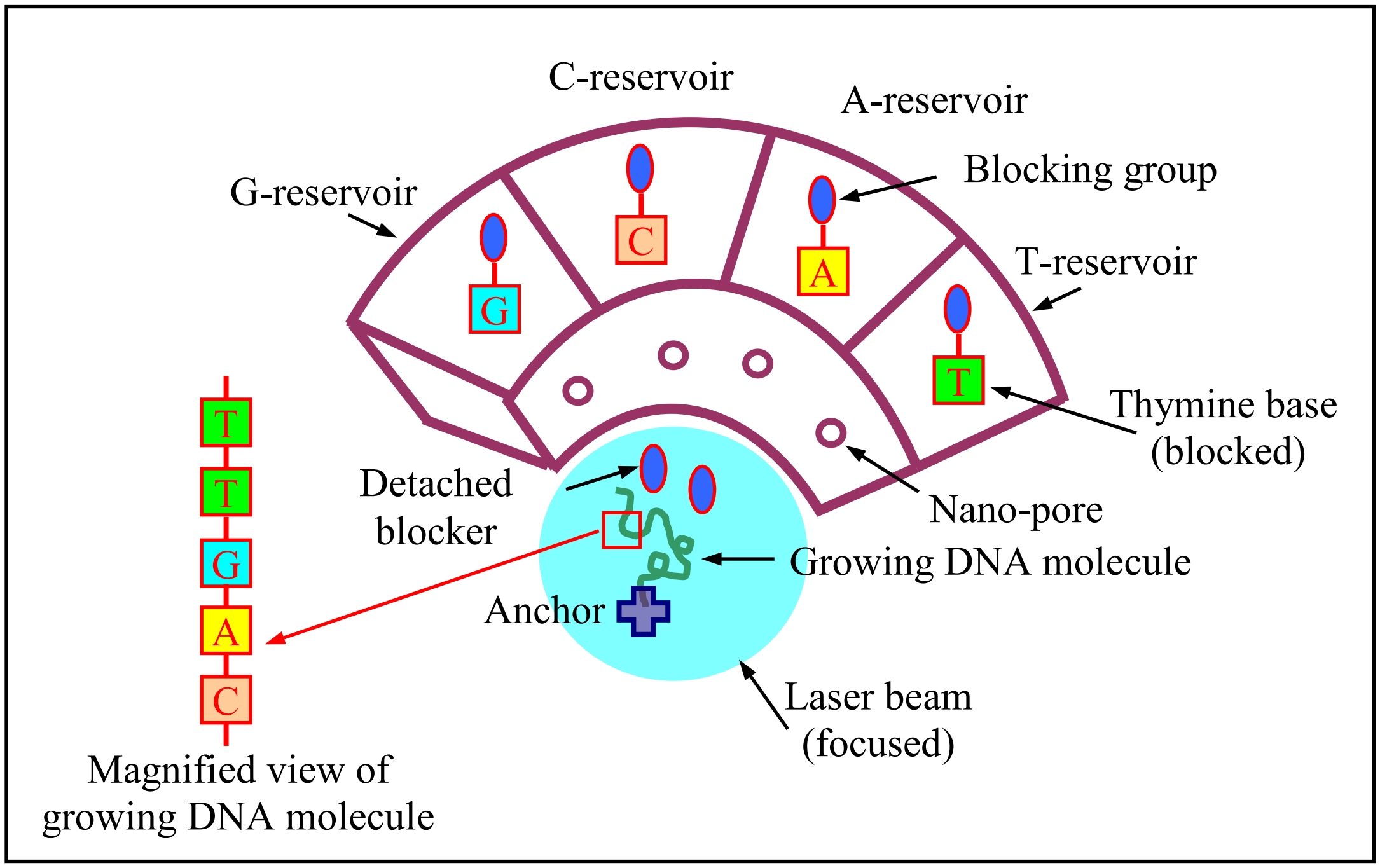

**Fig. 12**. Write station for creating DNA molecules of arbitrary base sequence. Four isolated chambers act as reservoirs for the four deoxyribonucleic acid bases; all base molecules stored in these reservoirs are blocked by a chemical group to ensure the attachment of a single base to the growing DNA strand at each stage of the process. Each reservoir is connected through a nano-pore to the main chamber, where a growing DNA molecule is anchored at one end but is free to float otherwise. When the growing DNA strand enters a reservoir through its nano-pore, it adds the desired base to its growing end, then returns to the main chamber. A photo-activation process in the main chamber (initiated by the focused laser beam) removes the blocker from the growing end of the strand. Upon completion of recording, the DNA molecule is detached from the anchor and transferred to a designated parking spot.

Whenever the activator is energized, the exposed active molecule is transformed from its native state to a (physically or chemically) different "excited" state. In Fig. 13(b) the two states of the active molecule are shown as differently colored. The native (ground) state and the transformed (excited) state of the active molecule must be stable; they must also be distinguishable from each other through the mechanism employed in the read station. (Once the entire strand is written, that is, when the active molecules are selectively converted from the ground state to the excited state, one may excise and remove the inert spacers while splicing the active molecules together without changing their sequential order.) The recorded strand thus created represents a binary sequence, whereby the ground state of the active molecule represents the binary digit "zero", and the excited state represents "one," or vice versa. If active molecules having more than two stable states are embedded in the precursor strand, non-binary (e.g., ternary, quaternary) recording will be possible as well.

The recorded strands will be "erasable" if the excited state(s) of the active molecule can somehow be reversed; otherwise the recorded strand will be an example of a write-once storage medium. In the latter case, however, erasable or rewritable data storage will still be possible in the following sense: any recorded strand that is no longer needed will be removed from its parking spot and destroyed (or abandoned), and a new precursor strand is written with fresh data, and stored in the same physical location (i.e., parking spot) from which the abandoned strand had been removed.

In this proposed scheme, the inert spacer molecules are needed only if the dimensions of the activator (i.e., the write head) are greater than those of the active molecule. In the case of an ultra-violet (UV) activator of wavelength $\lambda = 200$nm, for instance, if far-field optics are used to focus the beam, the diameter of the focused spot cannot be much smaller than 100nm. Since typical active molecules have dimensions on the order of 1 nm, the required spacer molecules must be at least 100 nm long. With near-field optics, it is possible to confine the optical activators to sub-wavelength dimensions, thereby reducing the required length of the spacer molecules. Electric-field or tunneling-tip activators may require even shorter spacers.

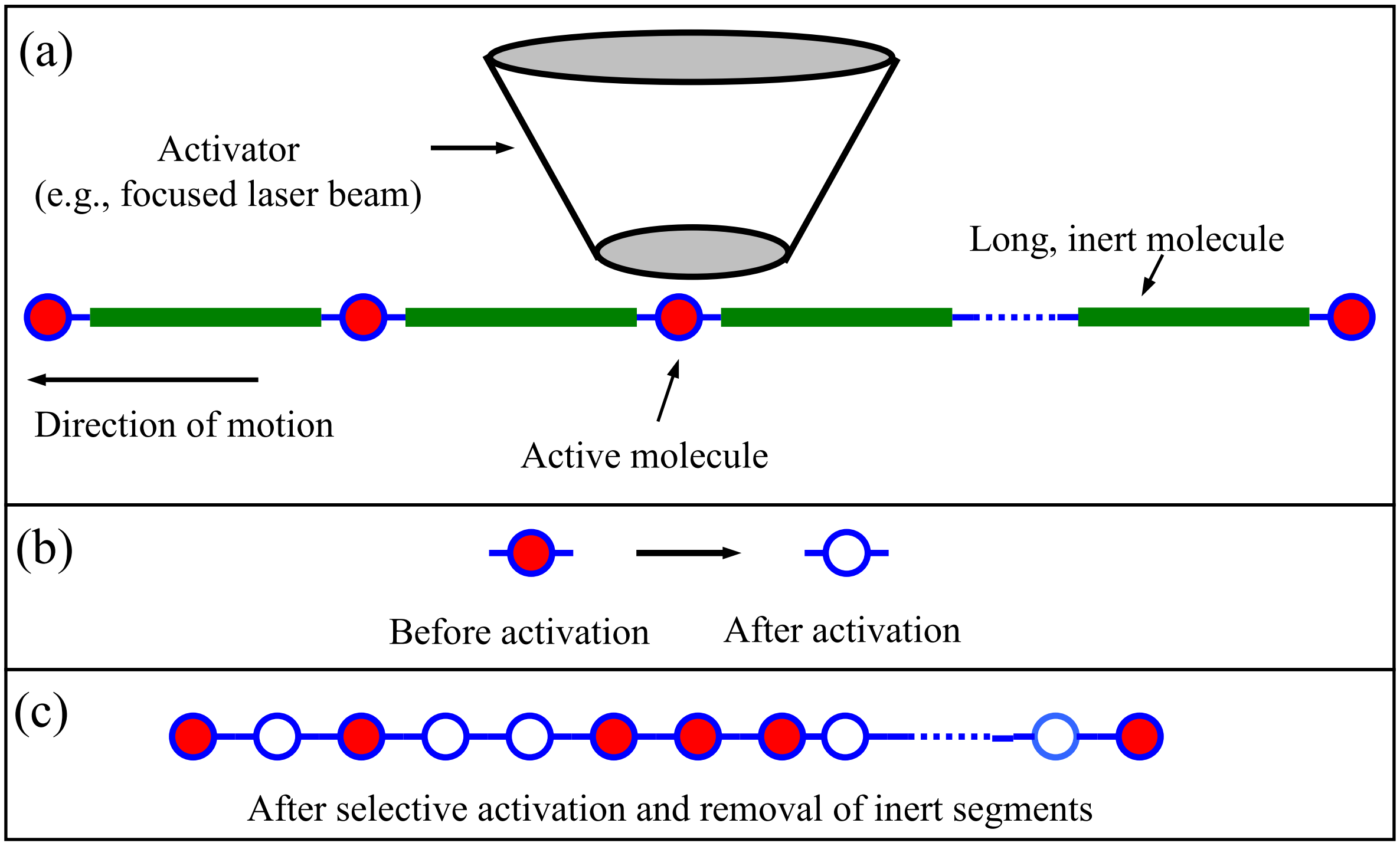

**Figure 13**. (a) A precursor strand consisting of long, inert segments separating active molecules in their native (or ground) state, is exposed to an activator. The strand is being pulled under the activator by a nano-actuator. (b) When the activator is energized, the active molecules are transformed from their native (ground) state to an excited state. If the activator is turned off, however, the active molecules remain in their ground state. (c) Once the entire precursor strand is written (i.e., its active molecules selectively placed in the excited state), the inert spacer molecules may be removed, and the active molecules spliced together in the same order in which they were arranged by the activator.

**Access Time**. An important characteristic of our data storage system, known as *access time*, is the average time it takes a macromolecular data block to travel from its parking spot to the read/write station. To simplify the analysis, let us assume that the macromolecule is bundled up in the form of a sphere of radius $r_0 = 0.5\ \mu\text{m} = 5 \times 10^{-7}$ m, moving at $V = 10$ m/s through water. At $T = 20°\text{C}$, water has density $\rho = 10^3$ kg/m$^3$ and viscosity $\eta = 10^{-3}$ Newtons.sec/m$^2$. The Reynolds number for the spherical particle moving through water would thus be $R = 2\rho V r_0 / \eta \approx 10$, which is fairly small.[5] This means that the flow of the liquid past the micron-sized sphere will be steady and laminar.

The drag force on the surface of the sphere will be $F = 6\pi\eta r_0 V = 10^{-7}$ Newtons.[5] Suppose the applied voltage across a 1 cm-wide chip is 10 V, so the electric field acting on the spherical particle will be $E = 10^3$ V/m. Since $F = qE$, the particle must have a charge of $q = 10^{-10}$C in order to move at the desired velocity under the applied field. Since a single-stranded DNA molecule has (on average) 2.5 electrons per base (i.e., $4 \times 10^{-19}$C per base), the macromolecule must have $\sim 2.5 \times 10^8$ nucleic acid bases. (This corresponds to a 60 Mbyte block of data associated with each molecule.) Since each base occupies roughly 1 nm$^3$, a $2.5 \times 10^8$-base molecule should fit in a micron-sized sphere, in agreement with our earlier assumptions.

The conclusion is that, if a polymer consisting of $10^8 - 10^9$ building blocks (monomers) is used to encode each data block, and if each monomer happens to have a few unpaired electrons, then the molecule can be transferred between its parking spot and the read/write station in a matter of milliseconds under the influence of a reasonable electric field (e.g., 10 Volts across a $1 \times 1$ cm$^2$ chip). We are aware, of course, that a charged polymer as assumed cannot be bundled up into a compact sphere, and a more realistic analysis needs to be undertaken in order to estimate the actual access time of the proposed system. Experimental data from gel electrophoresis indicate that the travel time of macromolecules under the above conditions should be on the order of a few seconds (rather than milliseconds). We plan to investigate this question further and, if necessary, consider alternative techniques for transporting our molecular data blocks. For instance, it is possible to neutralize the charge on a single-stranded DNA molecule by pairing it with its complement (i.e., creating double-stranded DNA), then place the uncharged molecule in a small lipid vesicle (similar to those used in biological systems) and move the vesicle across the system using either an electromagnetic field or a flow field created by the action of a micro-pump.

**Concluding remarks**. The objective of our research program is to build a data storage device using macromolecules, in general, and DNA molecules, in particular, as the storage medium. We use micro-fluidic techniques for transporting the molecules through various chambers (i.e., read, write, erase, and storage chambers) within a small, integrated chip. To date we have succeeded in building several micro-chambers, conducted DNA translocation experiments in different-sized chambers, and also demonstrated the principle of optical pumping in these devices. Future work includes fabrication of nano-pores in μ-chambers, conducting translocation experiments within these μ-chambers, and investigating various schemes of macromolecular writing and transportation.

**Acknowledgement**. The authors are grateful to Professors Raphael Gruener and Michael Hogan of the University of Arizona for illuminating discussions. The paper is based upon work supported in part by the *National Science Foundatio*n STC Program, under Agreement No. DMR-0120967.